\documentclass{emulateapj}
\usepackage{booktabs}

\let\url\relax
\usepackage[dvips]{hyperref}

\newcommand\thor[2]{$\theta^{#1}$Ori~#2}
\newcommand{\thC}{\thor{1}{C}}
\newcommand{\thA}{\thor{2}{A}}
\newcommand{\kms}{\ensuremath{\mathrm{km\ s}^{-1}}}
\newcommand{\pcc}{\ensuremath{\mathrm{cm}^{-3}}}

\newcounter{ionstage}
\renewcommand{\ion}[2]{\setcounter{ionstage}{#2}%
  \ensuremath{\mathrm{#1\,\scriptstyle\Roman{ionstage}}}}
\newcommand\OIlam{[\ion{O}{1}]\,6300\,\AA\@}
\newcommand\SIIlam{[\ion{S}{2}]\,6731\,\AA\@}

\newcommand\SIIIlam{[\ion{S}{3}]\,6312\,\AA\@}
\newcommand\NIIlam{[\ion{N}{2}]\,6584\,\AA\@}

\newcommand\OI{[\ion{O}{1}]}
\newcommand\SII{[\ion{S}{2}]}
\newcommand\SIII{[\ion{S}{3}]}
\newcommand\OIII{[\ion{O}{3}]}
\newcommand\NII{[\ion{N}{2}]}
\newcommand\HII{\ion{H}{2}}

\newcommand\HH[1]{HH~#1}
\newcommand\HHnew{\HH{873}}

\newcommand\OW[2]{#1--#2}
\newcommand\xy[2]{($#1$, $#2$)}

\newcommand\Vhel{\ensuremath{V_\odot}}
\newcommand\Vtan{\ensuremath{V_\mathrm{t}}}
\newcommand\eden{\ensuremath{n_\mathrm{e}}}
\newcommand\hour{\ensuremath{^\mathrm{h}}}
\newcommand\minute{\ensuremath{^\mathrm{m}}}

\def\adsurllinklabel{}

\makeatletter
\renewcommand{\fps@figure}{tp}
\makeatother

\setkeys{Gin}{width=\linewidth,height=0.9\textheight,keepaspectratio=true}

\begin{document}


\title{Velocity Structure in the Orion Nebula. I.~Spectral Mapping in
  Low-Ionization Lines}

\shorttitle{Spectral Mapping of the Orion Nebula}

\author{Ma.\ T. Garc\'{\i}a-D\'{\i}az\altaffilmark{1} and W. J. Henney}

\affil{Centro de Radioastronom\'{\i}a y Astrof\'{\i}sica, Universidad
  Nacional Aut\'onoma de M\'exico, Campus Morelia, Apartado
  Postal~3-72, 58090~Morelia, Michoac\'an, M\'exico}

\email{\url{tere@astrosen.unam.mx}, \url{w.henney@astrosmo.unam.mx}}

\altaffiltext{1}{Current address: Instituto de Astronom\'\i{}a,
  Universidad Nacional Aut\'onoma de M\'exico, Apartado Postal~877
  22800~Ensenada, Baja California, M\'exico}

\shortauthors{Garc\'{\i}a-D\'{\i}az \& Henney}

\begin{abstract}
  High-dispersion echelle spectroscopy in optical forbidden lines of
  O$^0$, S$^+$, and S$^{2+}$ is used to construct velocity-resolved
  images and electron density maps of the inner region of the Orion
  nebula with a resolution of $10~\kms{}\times3\arcsec \times
  2\arcsec$.  Among the objects and regions revealed in this
  study are (1)~the Diffuse Blue Layer: an extended layer of
  moderately blue-shifted, low-density, low-ionization emission in the
  southeast region of the nebula; (2)~the Red Bay: a region to the
  east of the Trapezium where the usual correlation between velocity
  and ionization potential is very weak, and where the emitting layer
  is very thick; and (3)~\HHnew{}: a new redshifted jet to the
  southwest of the Trapezium.

  \textbf{Note: A version of this paper with full-resolution figures
    can be obtained from
    \url{http://www.ifront.org/wiki/Spectral_Mapping_Paper_I}}
\end{abstract}


\journalinfo{Astronomical Journal, in press. March 2007}

\section{Introduction}
\label{sec:introduction}

The Orion Nebula is the prime example of a blister-type \HII{} region
\citep{1978A&A....70..769I}, in which ultraviolet radiation from the
high-mass Trapezium stars heats and ionizes the front surface of the
molecular cloud OMC-1, giving rise to a ``champagne flow'' of ionized
gas \citep{1979A&A....71...59T, 1986ARA&A..24...49Y, 2005ApJ...627..813H}
away from the cloud. Although the basic empirical model was proposed
more than 30 years ago \citep{1973ApJ...183..863Z}, the nebula
continues to present theoretical challenges due to the extraordinarily
rich phenomenology that is continually uncovered by advances in
observational techniques \citep[see][and references
therein]{2001ARA&A..39...99O}. The Orion Nebula presents one of the
best opportunities available for studying in detail the star formation
process in high density environments.  A rich cluster of more than a
thousand low-mass stars is associated with the nebula and centered on
the Trapezium \citep{1997AJ....113.1733H}. There is evidence that the
star formation rate has accelerated over the past few million years
\citep{1998ApJ...492..540H}, with the youngest stars concentrated
towards the center of the cluster. Indirect evidence from the survival
of the proplyds in the core of the nebula \citep{1999AJ....118.2350H}
indicates that the highest mass stars are probably even younger
($10^4$--$10^5$~years).  Dense molecular gas is primarily concentrated
in a north-south oriented lane \citep{1999ApJ...510L..49J},
displaced $\simeq 30\arcsec$ to the west of the Trapezium, containing
embedded sites at which star formation is still ongoing
\citep{2004ApJ...610L.117S}. A thorough understanding of the structure
and dynamics of the \HII{} region is necessary in order to understand
the role of the high-mass stars in limiting the star formation process
by dispersing the molecular cloud and in evaporating circumstellar
material around young low-mass stars \citep{2005ASPC..341..107H}.

Previous global studies of the nebula have been mainly based on
imaging in emission line filters \citep{1991ApJ...369L..75H,
  1992ApJ...399..147P} or spectrophotometry
\citep{1991ApJ...374..580B,1998MNRAS.295..401E}, and therefore only
give an average over different emitting regions along the line of
sight.  High-resolution spectroscopic studies show a wealth of detail
with many ions showing complex multi-component kinematic profiles
\citep{CampbellMoore1918, 1959ApJS....4..199W, 1973A&A....29..341D,
  1984A&A...137..245G, 1987A&A...176..347H, 1987ApJ...315L..55C,
  1992PhDT........35J, 1992ApJ...387..229O, 1993MNRAS.260..625M,
  1993MNRAS.262L..48M, 1993ApJ...403..678O, 1993ApJ...409..262W,
  1995MNRAS.273..615M, 2001ApJ...556..203O}. Although in principle
high resolution optical spectroscopy can allow the disentanglement of
different emission regions, in practice progress has been slow due
both to the complex nature of the object and the incomplete nature of
the data-sets available. Global spatio-spectral studies of the nebula
have generally been carried out with Fabry-Perot instruments
\citep{1997AJ....114.2016O, 2001AJ....122.1928R, 2003RMxAA..39..127D},
the velocity resolution of which ($>30~\kms{}$) is not sufficient to
study any but the highest velocity flows in the
nebula.\footnote{Earlier Fabry-Perot studies (e.g.,
  \citealp{1973A&A....29..341D}) were carried out at higher
  resolution, but employed photographic plates, thus making
  quantitative analysis difficult.} Radio recombination line studies
\citep[e.g.,][]{1997A&A...327.1177W} offer the possibility of higher
spectral resolution but have been confined to hydrogen lines, where
the thermal Doppler broadening of $\simeq 20~\kms$ largely negates
this advantage. On the other-hand, most multi-line high-resolution
($\sim 5~\kms{}$) echelle spectroscopy \citep{1999AJ....118.2350H} has
only covered restricted areas of the nebula.  A notable exception is
the recent study by \citet[hereafter DOH04]{2004AJ....127.3456D},
which covered the entire inner region of the nebula in the
[\ion{N}{2}], H$\alpha$ and [\ion{O}{3}] lines with a resolution of
$2\arcsec \times 1.5\arcsec \times 8$~km~s$^{-1}$.  In this paper, we
present a study in the lines of [\ion{S}{2}]$\lambda$6716,6731,
[\ion{O}{1}]$\lambda$6300, [\ion{S}{3}]$\lambda$6312, which is
complementary to the DOH04 study. Whereas DOH04 concentrated on the
high-velocity emission from jets and HH objects, this paper emphasizes
the more subtle variations in the bulk of the nebular gas, which
typically moves at transsonic velocities of 5--20~km~s$^{-1}$.

The four lines that we study cover a range of ionization stages from
low to moderate ionization. These correspond to different spatial
zones as a result of the ionization stratification that is typically
found in ionized nebulae. The [\ion{O}{1}] line is emitted principally
by the partially ionized gas in the ionization front, whereas the
[\ion{S}{3}] line is emitted only by ionized gas in the interior of
the nebula, where hydrogen is fully ionized, although excluding the
most highly ionized interior of the nebula, where sulfur is triply
ionized. The emission of the [\ion{S}{2}] lines contains contributions
from both the fully ionized and partially ionized zones and
furthermore the difference in critical density of the two lines of the
doublet allows the determination of the electron density. Our set of
lines is of generally lower ionization than that studied in DOH04
since the \SII{} and \OI{} lines extend to more neutral regions than
does \NII{}, whereas \SIII{} is weak from the most highly ionized
regions, which emit strongly in \OIII{}. Previous, low spectral
resolution studies of the electron density in the Orion nebula using
these lines have suffered from the problem that the derived densities
are merely an average of the conditions in different regions of the
nebula, weighted by the [\ion{S}{2}] emissivity. However, by
kinematically resolving the different components of the line, we are
able for the first time to separately determine the electron density
in different zones of the nebula.

Some preliminary results from this study have already been presented
in \citet{2005ApJ...627..813H}, where the patterns of global
kinematics in the nebula were compared with the results of
hydrodynamical simulations of champagne flows. The current paper
contains a much fuller description of our dataset and an empirical
analysis of the complex structure of the spatio-kinematic maps,
concentrating on velocities within $\pm30~\kms{}$ of the systemic
velocity and presenting evidence for the existence of many new nebular
features. Our results on the low-velocity \HH{528} outflow are
reported in a separate paper \citep{Henney-OrionSouth}, whereas
another companion paper (\citealp{Henney-Garcia-2006}, referred to
hereafter as Paper~II) will present a more detailed study of some of
the new features we have identified in the nebula, including
comparison with high-resolution \textit{HST} imaging.

In \S~\ref{sec:observations} we describe our observations and the data
reduction steps that we followed in order to produce flux- and
velocity-calibrated spectra and velocity channel maps of emission and
electron density. In \S~\ref{sec:veloc-struct-nebula} we use these
results to empirically derive the physical conditions and kinematics
of different features in the nebula, many of which have not been
previously discussed in the literature.

\section{The Observations and Data Reduction} 
\label{sec:observations}

\begin{figure*}
  \includegraphics{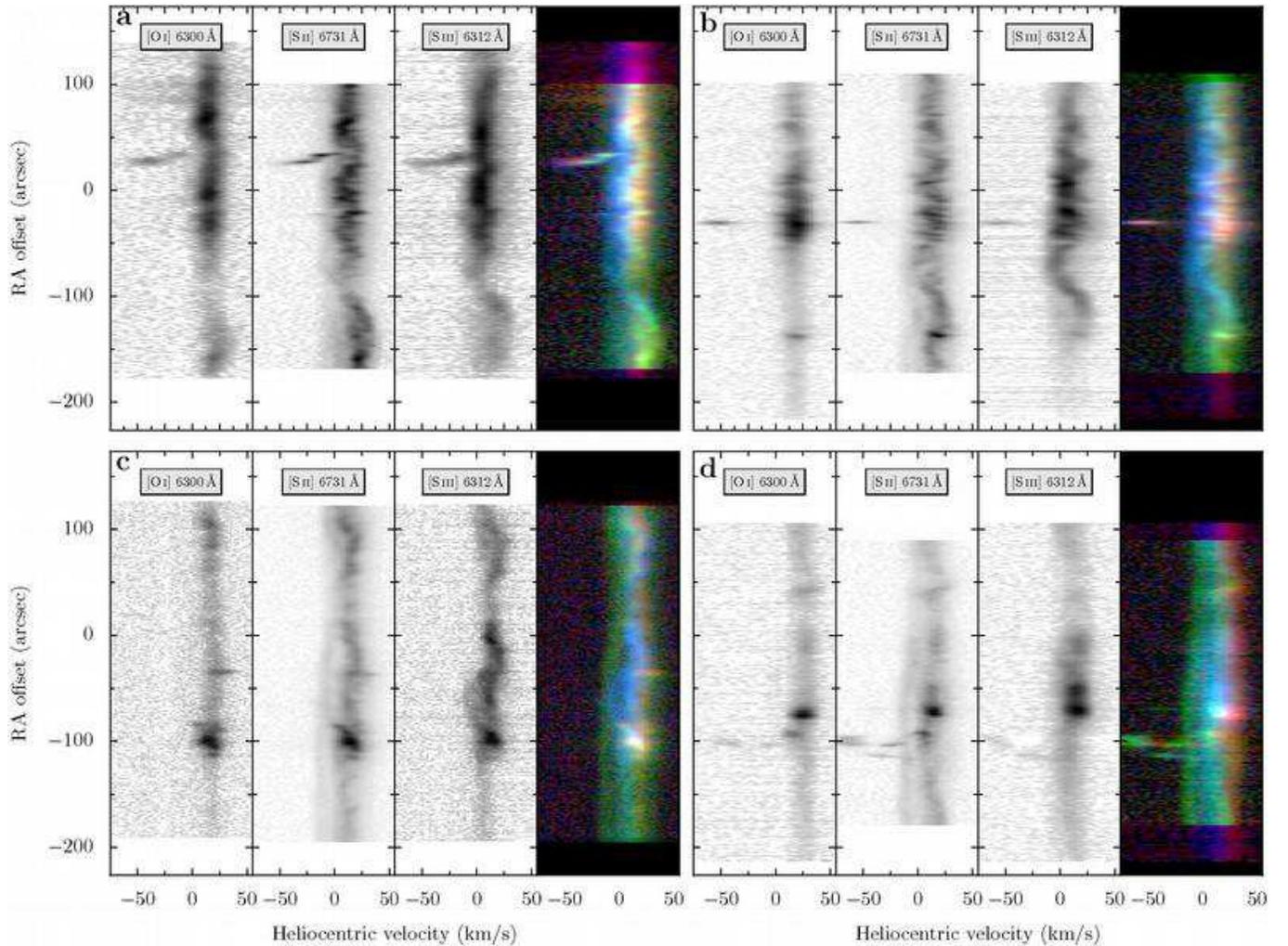} 
  \caption{Sample two-dimensional slit spectra of \OIlam{}, \SIIlam{},
    and \SIIIlam{} for four representative slit positions. All spatial
    offsets are measured with respect to \thC{}: (\textit{a})
    $68.8\arcsec$ W, (\textit{b}) $22.2\arcsec$ W, (\textit{c})
    $40.4\arcsec$ E, and (\textit{d}) $88.4\arcsec$ E. The grayscale
    negative images are each scaled to the maximum intensity in each
    slit and use a ``square root'' transfer function, which emphasizes
    fainter features.  The color, positive-scaled images are
    composites of the three emission lines for each slit, with \OI{}
    represented by red, \SII{} by green, and \SIII{} by blue.}
  \label{fig:bispectra}
\end{figure*}

\begin{figure*}
  \centering
  \includegraphics{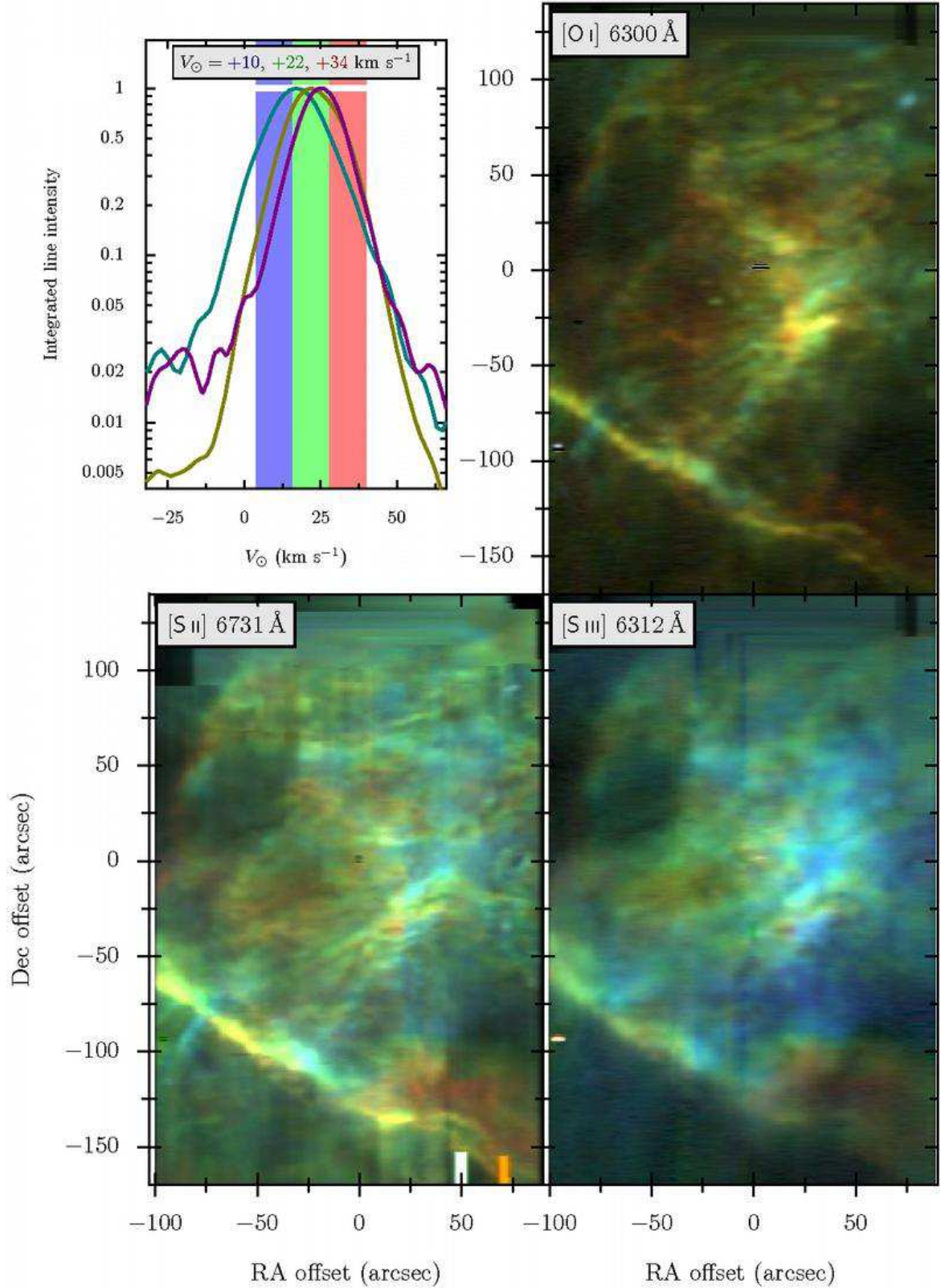} 
  \caption{Natural-weighted channel maps of the line cores,
    color-coded as shown in the top-left panel, showing the coarse
    velocity changes within $\pm 18~\kms$ of the peak \SII{} velocity
    of $\Vhel = +22~\kms$. For each emission line, an identical
    intensity scale is used for each of the three RGB channels,
    allowing meaningful comparison to be made between the colors in
    the different maps. The upper-left panel indicates the position of
    each velocity channel on the integrated spectrum of the entire
    nebula (logarithmic scale) in the three lines. Left to right:
    \SIII{} (turquoise), \SII{} (gold), \OI{} (purple).}
  \label{fig:channel-broad}
\end{figure*}

\begin{figure*}
  \centering
  \includegraphics{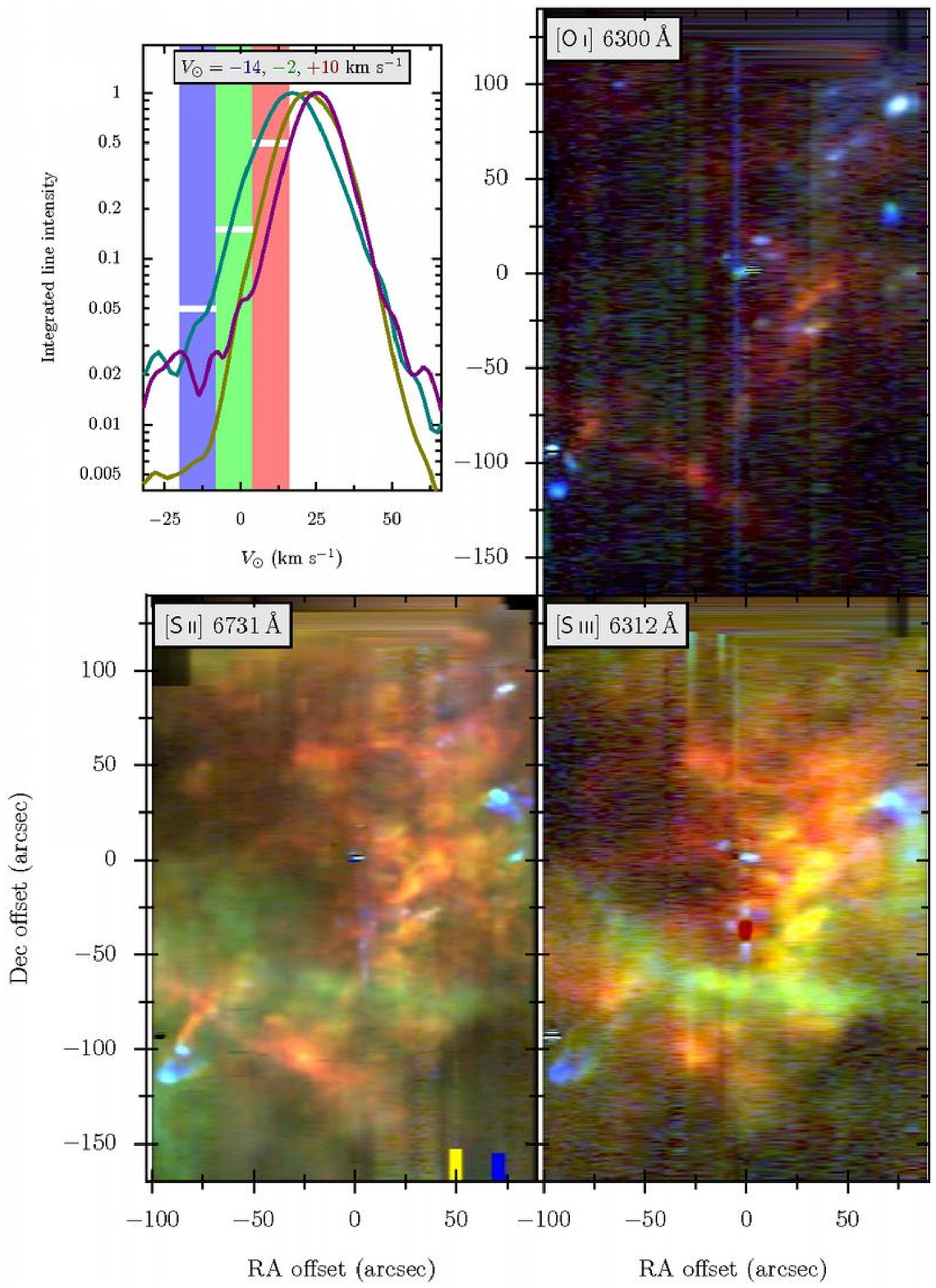} 
  \caption{Same as Fig.~\ref{fig:channel-broad} but for the blue
    side of the line profiles, corresponding to heliocentric
    velocities between $\Vhel = -20~\kms$ and $+16~\kms$. Since the
    emission line profiles are falling towards the blue in this
    velocity range, different normalizations are used for the images
    in the 3 velocity bands, as indicated by the horizontal white
    lines in the top-left panel.}
  \label{fig:channel-broad-blue}
\end{figure*}
\begin{figure*}
  \centering
  \includegraphics{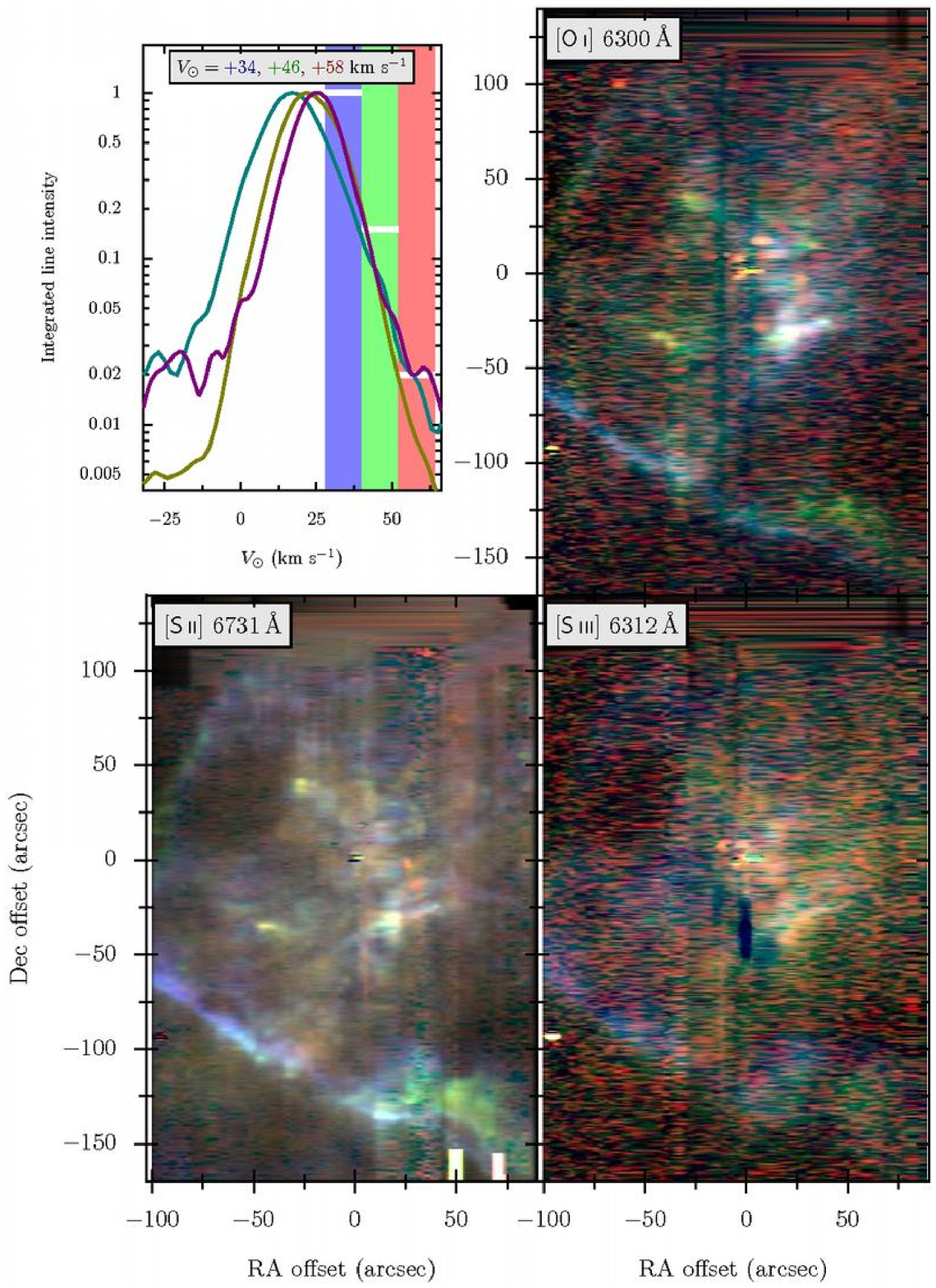} 
  \caption{Same as Fig.~\ref{fig:channel-broad} but for the red
    side of the line profiles, corresponding to heliocentric
    velocities between $\Vhel = +28~\kms$ and $+64~\kms$. Since the
    emission line profiles are falling towards the red in this
    velocity range, different normalizations are used for the images
    in the 3 velocity bands, as indicated by the horizontal white
    lines in the top-left panel.}
  \label{fig:channel-broad-red}
\end{figure*}

High-resolution spectroscopic observations of the Orion Nebula were
obtained from two different observatories: one set from Kitt Peak
National Observatory (KPNO), and the other set from Observatorio
Astron\'omico Nacional at San Pedro M\'artir, B.C., Mexico (SPM). The
first set of spectra was obtained with the echelle spectrograph
mounted on the 4~m telescope at KPNO and is described in detail in
\citet{2001ApJ...556..203O} and DOH04. The $300 \times 0.8''$ slit was
oriented north-south on the nebula and exposures were taken at a grid
of different declinations (separation $\approx 2''$).  This dataset,
with velocity resolution of 8~\kms{}, consists of 37 positions for
[\ion{S}{2}]$\lambda\lambda$6716,6731 \AA{} lines in two disjoint
regions of the nebula: one in the East, centered on $\mathrm{RA} =
05^\mathrm{h}35^\mathrm{m}21\fs5$, and one in the West, centered on
$\mathrm{RA} = 05^\mathrm{h}35^\mathrm{m}11\fs70$.

The SPM observations were obtained in 2002 October and 2003 January
for [\ion{S}{2}] (55 pointings) and in December 2003 for the
[\ion{O}{1}] and [\ion{S}{3}] lines (60 pointings) using the MES-SPM
instrument \citep[Manchester Echelle
Spectrometer,][]{2003RMxAA..39..185M} attached to the 2.1~m telescope
in its f/7.5 configuration. For the majority of the exposures, we used
a 150$\mu$m ($\equiv$ 2 arcsec) slit, giving a velocity resolution
(FWHM) of 12~\kms{}, while for some exposures we used a 70$\mu$m
($\equiv$ 0.95 arcsec) slit, giving a velocity resolution of
6~\kms{}. The $312\arcsec$-long slit was oriented North-South for most
of the observations, except for one EW-oriented exposure in each line
at a position $9.6\arcsec$ south of \thC{}{}. We used the star JW499
\citep[$\alpha = 05^\mathrm{h}35^\mathrm{m}15\fs89$, $\delta =
-05\arcdeg 23\arcmin 49\farcs92$;][]{1988AJ.....95.1755J} as a
reference point for all of our pointings.  For each position we took
two spectral exposures with duration of 450~s for the 150~$\mu$m slit
(with $2\times 2$ pixel on-chip binning), or 900~s for the 70~$\mu$m
slit (with no on-chip binning).  Thorium-Argon lamp spectra were taken
for wavelength calibration between each slit position.  In order to
establish the exact position of the slit in each pointing we took
direct ``slit images'' of short duration, in which the diffraction
grating was replaced by a mirror, with the slit removed for half of
the duration of the exposure. This was necessary since the telescope
pointing proved insufficiently reliable to guarantee $2\arcsec$
offsets between exposures.\footnote{This temporary pointing problem
  at the SPM observatory has since been solved.}


Combining the datasets from the two observatories gives 92 NS
pointings in the [\ion{S}{2}] doublet, spanning an interval of
190.6$''$ in RA\@.  For the [\ion{O}{1}] and [\ion{S}{3}] lines, which
were observed simultaneously due to their proximity in wavelength, we
have only the 60 NS pointings from SPM, spanning a similar interval in
RA\@.

The initial stages of data reduction (bias removal, flat fielding,
cosmic ray removal) were carried out using standard IRAF\footnote{IRAF
  is distributed by the National Optical Astronomy Observatories,
  which is operated by the Association of Universities for Research in
  Astronomy, Inc. under cooperative agreement with the National
  Science foundation} tasks. This was followed by rectification and
first-order wavelength calibration of the two-dimensional spectra
based on the comparison lamp spectra, also using standard IRAF tasks.

After transforming all the spectra to a common heliocentric velocity
frame, we carried out a series of further steps in order to produce
internally consistent, well-calibrated velocity cubes in each of the
four emission lines. These were implemented using purpose-written
Fortran and Python routines.
\begin{enumerate}
\item Continuum emission was removed by fitting a quadratic function
  to each row of each two-dimensional spectrum, including only
  line-free regions in the fit.
\item An astrometric solution was found for each of the ``slit
  images'' mentioned above, using several of the Trapezium cluster
  stars. This allowed us to accurately determine the slit position of
  each exposure.
\item Due to variations in atmospheric transparency between exposures,
  it was necessary to perform a slit-to-slit relative intensity
  calibration. This was done by comparison with the spectrum from the
  horizontal slit position.
\item In the same way, the relative slit-to-slit wavelength
  calibration was refined using the horizontal slit spectrum. The line
  centroid velocity as a function of position was calculated for each
  slit and small shifts were introduced in the velocity scale to force
  agreement with the horizontal slit at the point where they
  crossed. The shifts required were typically less than 2~\kms{}.
\item For the [\ion{O}{1}] line it was necessary to remove the
  night-sky component, which occurs at 0~\kms{} in the geocentric
  frame. We had chosen the dates of the observations such that the
  heliocentric correction means that this component is not blended
  with the main emission from the nebula. Since the profile of this
  component was seen to be constant along the length of the slit, it
  was straightforward to remove by fitting an analytic profile to the
  summed spectrum of each slit (a hyper-Gaussian profile was
  used). The centroid of the fitted profile provided an additional
  check on the accuracy of the wavelength calibration for this line.
\end{enumerate}

Examples of the resultant two-dimensional spectra after following all
these steps are shown in Figure~\ref{fig:bispectra}. Only the longer
wavelength component of the [\ion{S}{2}] doublet is shown. In this and
all following figures, heliocentric velocities, $\Vhel$, are
used\footnote{The transformation to ``Local Standard of Rest''
  velocities is $V_\mathrm{lsr} = \Vhel - 18.1~\kms{}$.} and positions
are specified as arcsecond offsets ($x$, $y$) with respect to the
principal ionizing star of the nebula, \thC{}, which has J2000
coordinates $\alpha = 05\ 35\ 16.46$, $\delta = -05\ 23\ 23.2$
\citep{2000A&A...355L..27H}. Transformations between these offsets and
the commonly used coordinate-based naming scheme of
\citet{1992ApJ...387..229O} are given in the Appendix.

The preceding steps have not significantly degraded the spatial or
velocity resolution of the original data. However, in order to combine
the individual slit spectra into a three-dimensional data cube, it is
necessary to degrade the resolution to that of the lowest common
denominator. To that end, we carried out the following steps:
\begin{enumerate}\setcounter{enumi}{5}
\item The instrumental linewidths were homogenized by smoothing the
  SPM 70~$\mu$m spectra by 10~\kms{} and the KPNO spectra by 9~\kms{}
  to give the same instrumental width as for the SPM 150~$\mu$m
  spectra (12~\kms{} FWHM).
\item The slit spectra were binned/interpolated\footnote{We used an
    algorithm that reduces to binning in the case where the spacing
    between the original data is finer than the target grid spacing
    and reduces to linear interpolation in the case where it is
    coarser. The algorithm also takes care of interpolating over ``bad
    regions'' in certain spectra, which were due, for instance, to
    spectrograph reflections when the slit passes over a very bright
    star.} onto a uniform grid in RA, declination, and heliocentric
  velocity, with voxel size of $0.6\arcsec{} \times 0.6\arcsec{}
  \times 4~\kms{}$, in order to produce the three-dimensional data
  cube. During this stage, we also applied a ``de-jittering''
  procedure to each two-dimensional iso-velocity cut in order to
  eliminate unsightly striping due to residual inter-slit variations.
  For each velocity value, a mean intensity of extended diffuse
  emission\footnote{Defined as pixels within $3\sigma$ of the mean.}
  was calculated for each slit, and these were then filtered across
  the slits by a 5-point smoothing kernel to produce a correction to
  the intensity of each slit.
\end{enumerate}

\newcommand\D{\discretionary{}{}{}}
The resulting data cubes for the lines of the three ions are shown in
Figures~\ref{fig:channel-broad} to \ref{fig:channel-broad-red} as
velocity channel maps.\footnote{The data will be made publicly
  available via the ftp server
  \url{ftp://orion.phy.vanderbilt.edu/outgoing/ResearchScientist/OrionNebula}}
Each of the images encodes the emission in three adjacent 12~\kms{}
velocity channels as blue, green, and red, respectively, moving from
more negative to more positive velocities, as shown on the
logarithmically scaled line profile of the final panel. We estimate
the spatial resolution of the resulting images to be $3\arcsec \times
2\arcsec$ for \SII{} and $4\arcsec \times 2\arcsec$ for \OI{} and
\SIII{}. Despite our best efforts, some artefacts remain from the
combination of different datasets. These are identifiable as bright or
dark vertical stripes on the images, which are more prominent in the
line wings, where the intrinsic signal is weaker. They are chiefly
caused by imperfections in the rectification and continuum removal, or
in some cases due to reflections caused by a bright star falling in
the slit (particularly at an RA offset of $0$).

Further steps were necessary in order to photometrically calibrate our
data. We used spectrophotometric observations from the literature in
order to ``pin down'' the absolute line fluxes at different points of
the nebula by comparing with the velocity-integrated intensity that we
observe at the same point. For [\ion{O}{1}] and [\ion{S}{3}], we used the
observations of \citet{1991ApJ...374..580B}, who give the absolute
flux of the blend of these two lines, together with the higher
spectral resolution observations of \citet{2000ApJS..129..229B}, who
give the ratio of the two lines. In both cases the observations are
for an EW slit, positioned to the W of \thC{}, which overlaps with our
own data over a length of $\simeq 40\arcsec$. Note that, although we
obtain the [\ion{O}{1}] and [\ion{S}{3}] lines in the same spectrum,
we still need to externally calibrate their ratio because of the
unknown variation in the spectrograph efficiency over the wavelength
range of the two lines. Since we performed all the observations with
the same spectrograph settings, we have assumed that the relative
efficiency is the same for all slits, and, indeed, comparison with the
\citeauthor{2000ApJS..129..229B} data confirms this assumption.

For the [\ion{S}{2}] doublet, a greater variety of spectrophotometric
data exists in the literature \citep{1991ApJ...374..580B,
  1992ApJ...399..147P, 1998MNRAS.295..401E}. However, in this case, we
have the added complication that our data are a combination of results
from two different instruments at two different observatories. Hence
the calibrations of the SPM-MES and KPNO slits were performed
separately. The density-sensitive ratio of the two [\ion{S}{2}] lines
is of key importance, so we took particular care to verify our
velocity-integrated results against those of previous authors.

\begin{figure*}\centering
  \includegraphics{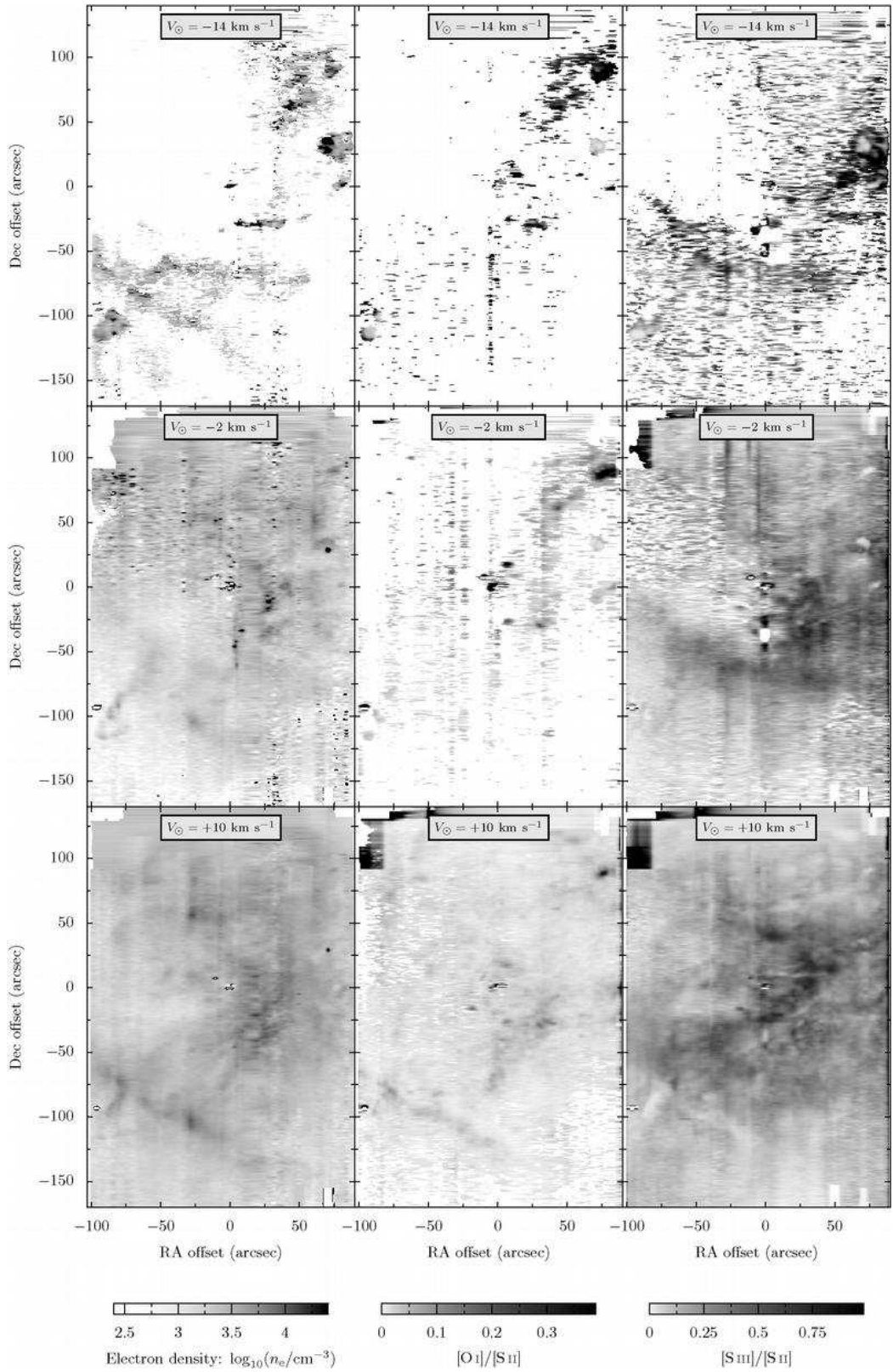} 
  \caption{Blueshifted isovelocity channel maps of \SII{}-derived
    electron density (logarithmic negative grayscale), \OI{}/\SII{}
    ratio, and \SIII{}/\SII{} ratio (linear negative grayscales). Each
    velocity channel is of width 12~\kms{}.  }
  \label{fig:ratio-a}		
\end{figure*}

\begin{figure*}\centering
  \includegraphics{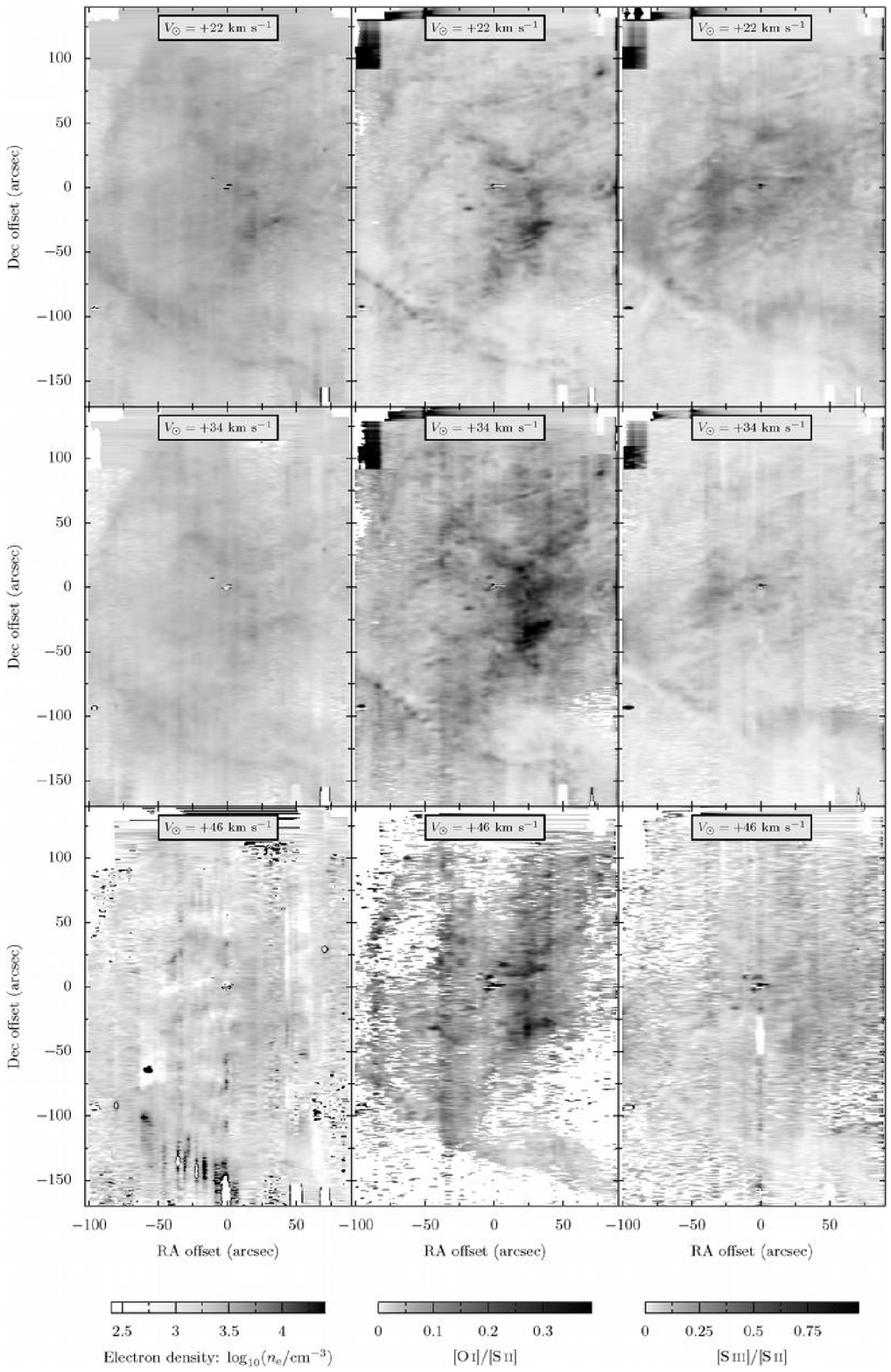}          
  \caption{Same as Figure~\ref{fig:ratio-a} but for redshifted
    velocities. }
  \label{fig:ratio-b}		
\end{figure*}

We have also calculated channel maps of electron density and of the
line ratios [\ion{O}{1}]/[\ion{S}{2}] and [\ion{S}{3}]/[\ion{S}{2}],
which are shown in Figure~\ref{fig:ratio-a} and~\ref{fig:ratio-b}. In
both cases, the sum of the two [\ion{S}{2}] lines was used in the
denominator and for the ratios with \OI{} and \SIII{} the \SII{} maps
were first smoothed slightly in order to better match the spatial
resolutions.


\section{Kinematic-ionization structure of the nebula}
\label{sec:veloc-struct-nebula}

\begin{figure*}\centering
  \includegraphics[height=0.7\textheight]{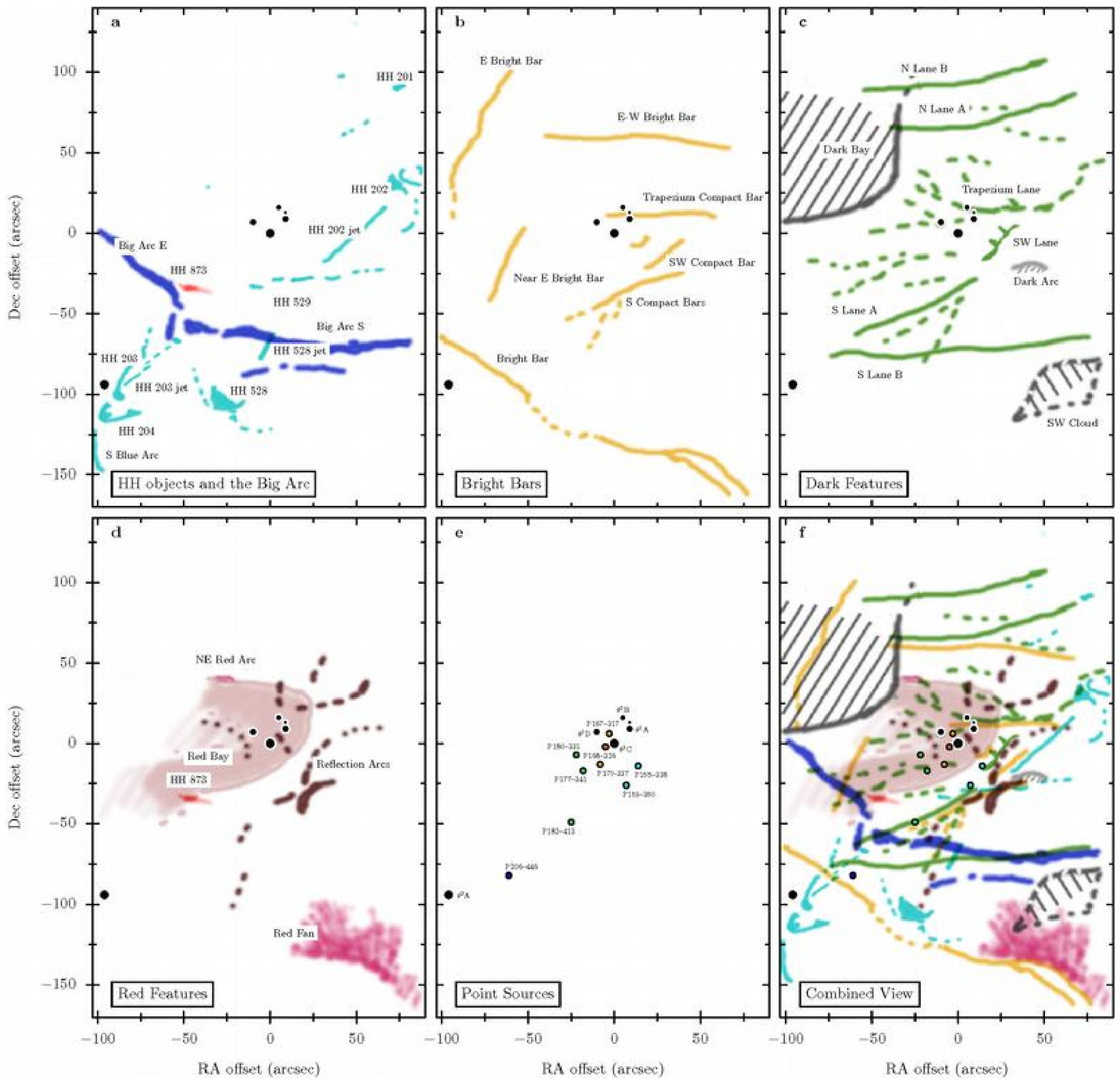} 
  \caption{Finding charts for features in our channel
    maps. (\textit{a})~Blueshifted features assosciated with the Big
    Arc (dark blue) and HH outflows (light blue), together with the
    newly identified redshifted jet (red). (\textit{b})~Bright Bars
    and newly identified compact bars (yellow). (\textit{c})~Dark
    regions corresponding to low-ionization lanes (olive green),
    foreground extinction features (mid gray), and the dark arc (light
    gray).  (\textit{d})~Redshifted features seen in higher-ionization
    lines (light pink) and lower-ionization lines (magenta), together
    with reflection filaments seen in the far-red
    (brown). (\textit{e})~OB stars (filled black circles) and proplyds
    (open circles). The shading of the proplyds indicates their peak
    \OI{} velocity from 20~\kms{} (blue) to 31~\kms{}
    (red). (\textit{f})~Superposition of all the features.}
  \label{fig:finding}
\end{figure*}

In this section we provide an empirical description of the nebular
structure and kinematics as revealed by our spectra. We begin with a
general overview of the kinematics and follow with more detailed
treatments of some of the individual objects and spectral features,
both those that are already known and those that are described for the
first time in this paper. We have made use of both the isovelocity
channel maps and the position-velocity spectra when identifying
features. The channel maps are most useful for taking advantage of the
image-processing capacity of the human brain in order to reveal
large-scale, spatially coherent features that are not readily apparent
in the individual slit spectra. However, it is always necessary to
return to the original position-velocity spectra, which in many cases
are of higher spectral resolution, in order to confirm identifications
and make precise measurements. Figure~\ref{fig:finding} illustrates in
cartoon form the location of all the features discussed below.

When naming new objects, we have tried to follow the established
conventions of the literature and to employ names that evoke only the
two-dimensional appearance of the feature on the plane of the sky. By
deliberately avoiding descriptions that presuppose a certain model of
the three-dimensional structure of the nebula, we hope to avoid the
problem of the names becoming obsolete with advances in our physical
understanding. Some widely used examples of the existing nomenclature
are \textit{rays}---narrow, precisely linear features, which may be
bright or dark, usually aligned with some source of illumination;
\textit{bars}---bright, approximately linear features;
\textit{lanes}---dark, approximately linear features;
\textit{arcs}---curved, more or less elongated features, which may be
dark or bright.

\subsection{Large-scale nebular properties}
\label{sec:glob-nebul-char}

\subsubsection{The line core}
\label{sec:line-core}

The general structure of the emission near the systemic velocity of
the three lines is shown in the 3-color isovelocity images of
Figure~\ref{fig:channel-broad}. Note that the centroid velocity of the
molecular gas behind the nebula is around $\Vhel = +28~\kms{}$, lying
between the green and the red channels of this figure. The emission is
very clumpy and filamentary in \OI{}, rather less so in \SII{}, and
much smoother and more diffuse in \SIII{}. The principal features of
the nebula are well known---see \citet{1982occs.book.....G} for a
compilation of early work on the subject. 
\begin{enumerate}
\item The region of peak nebula brightness, lying just to the
  southwest of the Trapezium, centered around \xy{25}{-25}. This is a
  region of highly inhomogeneous brightness, especially in the lower
  ionization lines, with much structure that is only poorly resolved
  in our observations, but that can be appreciated in \textit{HST}
  images.
\item The Bright Bar (e.g., \cite{1974Ap&SS..28..351E}), a prominent
  low-ionization, approximately linear emission feature to the
  southeast, running from \xy{-100}{-60} to \xy{90}{-170}. This is the
  ionized portion of the famous Orion Bar \citep{1976ApJ...204..420W},
  which is a concentration of dense molecular gas, offset to the
  southeast from the ionized emission. The region south of the Bright
  Bar is brighter in \SII{} than in either of the other lines.
\item The Dark Bay, a foreground extinction feature, which can be
  appreciated as a general reduction in brightness to the northeast,
  with a southern border running from \xy{-100}{0} to \xy{-30}{20}.
\end{enumerate}

As one passes from \OI{}, through \SII{}, to \SIII{}, the line core
becomes more weighted towards the blue overall, as can be appreciated
both from the integrated line profiles (shown in the top-left panel of
the figure) and from the general hue of each image.\footnote{In all of
  Figs.~\ref{fig:channel-broad} to \ref{fig:channel-broad-red}, the
  relative intensity scales of the 3 velocity channels is identical
  between the 3 lines, \OI{}, \SII{}, and \SIII{}, allowing meaningful
  color comparisons to be made between the lines.} However, this trend
is not universal across the face of the nebula and two broad regions
of redshifted emission are present in the \SIII{} map (shown in
fig.~\ref{fig:finding}\textit{d}. The first of these lies to the east
of the Trapezium, and we name this region the Red Bay since its
general shape seems to mimic that of the Dark Bay that lies to the
northeast of it. It is discussed further in \S~\ref{sec:red-bay}
below. The second region lies just north of the western end of the
Bright Bar, and we name it the Red Fan due to its appearance in low
ionization lines as a set of bright plumes that fan away from that
portion of the Bright Bar.

There is considerable agreement in the pattern of velocity changes
between the three different lines. In each case, the most blueshifted
gas is found in a broad C-shaped arc to the north, west, and south of
the Trapezium, which wraps around the Red Bay. Apart from the two
redshifted regions mentioned above, reduced blueshifts are also seen
in the Bright Bar and in various other large-scale filamentary
structures that criss-cross the nebula. In addition to the bright bars
that have previously been reported in the literature
\citep{2000AJ....120..382O}, we have identified several examples of
more compact bars towards the core of the nebula, which are labelled
in Figure~\ref{fig:finding}\textit{b}. Furthermore, we have also
detected a series of dark lanes, which are labelled in
Figure~\ref{fig:finding}\textit{c}. Owing to the narrowness of most of
these features, a full investigation of their nature requires a
detailed comparison with high-resolution \textit{HST} imaging, which
is beyond the scope of the current paper, but is carried out in
Paper~II.\@

\subsubsection{The blue flank}
\label{sec:blue-flank}
Figure~\ref{fig:channel-broad-blue} shows further isovelocity images,
but this time for the blueshifted flank of the emission lines. There
is an overlap with Figure~\ref{fig:channel-broad} in the sense that
the red channel of Figure~\ref{fig:channel-broad-blue} is the same as
the blue channel of Figure~\ref{fig:channel-broad}. All of the lines
are falling away sharply towards negative velocities in this velocity
range, so that, if a naturally weighted brightness scale (as in
Fig.~\ref{fig:channel-broad}) had been used, then all the images would
be dominated by the red channel. Instead, different scales are used
for the 3 channels (with maxima as indicated by white horizontal lines
in the top-left panel) so as to give a more balanced color scheme. As
a result, care must be taken in interpreting the images since the
brightest blue features are 10 times fainter than the brightest red
features.

The \OI{} line is generally very faint in these velocity channels, and
concentrated in very localized regions. In the most negative (blue)
channel, one sees only emission from high-velocity Herbig-Haro shocks:
\HH{201} to the northwest, \HH{202} to the west, \HH{269/507} to the
southwest and \HH{203/204} to the southeast. In the least negative
(red) channel, one sees both low-velocity Herbig-Haro objects, such as
\HH{528} to the south, and the blue wings of the bright bars.

In the \SII{} and \SIII{} lines, the most negative channel is again
dominated by the HH objects, but the other two channels show much more
extensive emission. In particular, the green channel, centered around
$\Vhel = -2~\kms$ is very prominent in the south of the nebula in
these two lines. Part of this emission is due to the Big Arc
\citep{1997AJ....114.2016O, 2004AJ....127.3456D}, a broad, elongated
swathe of intermediate velocity blueshifted emission, running from
southwest from \xy{-100}{0} to \xy{-50}{-60} and then more nearly west
to \xy{80}{-75} (see Fig.~\ref{fig:finding}\textit{a}). Although it is
visible in \SII{} (faintly) and \SIII{}, the Big Arc is most prominent
in higher ionization lines, such as \OIII{}, and it is discussed in
greater detail in Paper~II\@. In addition to the Big Arc, widespread
diffuse emission is seen in this velocity channel, being especially
strong in \SII{}. The \SII{} density sensitive line ratio map
(Fig.~\ref{fig:ratio-a}) show this emission to come from low-density
gas, which we refer to as the Southeast Diffuse Blue Layer and which
is discussed further in \S~\ref{sec:diffuse-blue-layer} below.

\subsubsection{The red flank}
\label{sec:red-flank}
Figure~\ref{fig:channel-broad-red} shows isovelocity images for the
redshifted flank of the emission lines, with the blue channel here
being identical to the red channel of
Figure~\ref{fig:channel-broad}. Again, unequally weighted brightness
scales have been used to compensate for the general fall-off in the
line profiles away from their core. This time, the brightest red
features are a factor of 50 times fainter than the brightest blue
features.

Even with this help, for \OI{} very little emission can be discerned
above the noise in the redmost channel ($\Vhel = +52$ to $+64~\kms$),
except for a faint glow in the extreme northwest. In the \SII{} and
\SIII{} images, on the other hand, faint but extensive emission is
found, concentrated towards the region southwest of the Trapezium. In
the \SII{} image, which has a significantly higher signal-to-noise
than the other two, the emission can be seen to be concentrated into
clumps and arcs (see Fig.~\ref{fig:finding}\textit{d}), which usually
do not coincide with the bright filaments seen at bluer velocities. We
believe that this emission arises from the scattering of nebular
emission lines by dust particles that lie behind the nebula. See
further discussion below in \S~\ref{sec:red-scatt-comp}.

Most of the jets and outflows in the Orion Nebula show only
blueshifted emission \citep[see][]{Henney-OrionSouth}, but we have
detected a hitherto unreported redshifted jet in our data, which is
clearly visible in the \OI{} and \SII{} images of
Figure~\ref{fig:channel-broad-red} at a position of
\xy{-40}{-35}. This is described in detail in \S~\ref{sec:hhnew}
below.

At the more moderate redshift of the blue and green channels ($\Vhel =
+28$ to $+52~\kms$), one sees mainly the red wings of the flows from
the bright bars (see Paper~II).

\subsection{Herbig-Haro outflows}
\label{sec:other-outflows}
The classical M42 Herbig-Haro objects \HH{201--204} and accompanying
jets \citep{1997AJ....114..730O} are all present in our spectra,
together with more recently recognised outflows such as \HH{529} (see
Fig.~\ref{fig:channel-broad-blue}). Although these flows do appear in
the velocity channels that we are considering, the majority of their
emission is at higher blueshifted velocities (see, for example,
Fig.~\ref{fig:bispectra}\textit{d}, which shows the \HH{203} jet and
the \HH{203/204} bowshocks). We therefore defer detailed discussion of
these objects to a later paper and merely note their presence here.

Other Herbig-Haro flows do not show such high velocities and the
entirety of their emission is contained within the velocity range
considered in this paper. These include the large, low velocity
\HH{528} flow and two objects that are described here for the first
time: \HHnew{}, a newly discovered redshifted jet, and the South Blue
Arc, which may represent a leading bowshock of the \HH{203/204}
system.

\subsubsection{\HH{528}}
\label{sec:hh528}

\HH{528} is a large-scale, low-ionization, low-velocity outflow in the
core of the Orion Nebula \citep{1999oisc.conf...25O,
  2000AJ....119.2919B}.  Although it is one of the largest outflows in
the nebula, its existence was recognized only relatively recently,
owing both to its lack of high-velocity emission and the fact that its
bowshock, despite being very large and bright, is superimposed on the
low-ionization emission from the Bright Bar. Previous spectroscopic
studies have been unable to kinematically resolve the outflow from the
systemic nebular emission. The multi-line velocity mapping that we
present here allows the emission from the outflow and its bowshock to
be cleanly separated for the first time from the more redshifted
nebular gas, showing it to be blueshifted by $10 \pm 3~\kms$ with
respect to the systemic velocity of the molecular cloud. Further
details of our radial velocity measurements of this object are given
in a separate paper \citep{Henney-OrionSouth}.

\subsubsection{\HHnew}
\label{sec:hhnew}

\begin{figure*}
  \centering
  \includegraphics{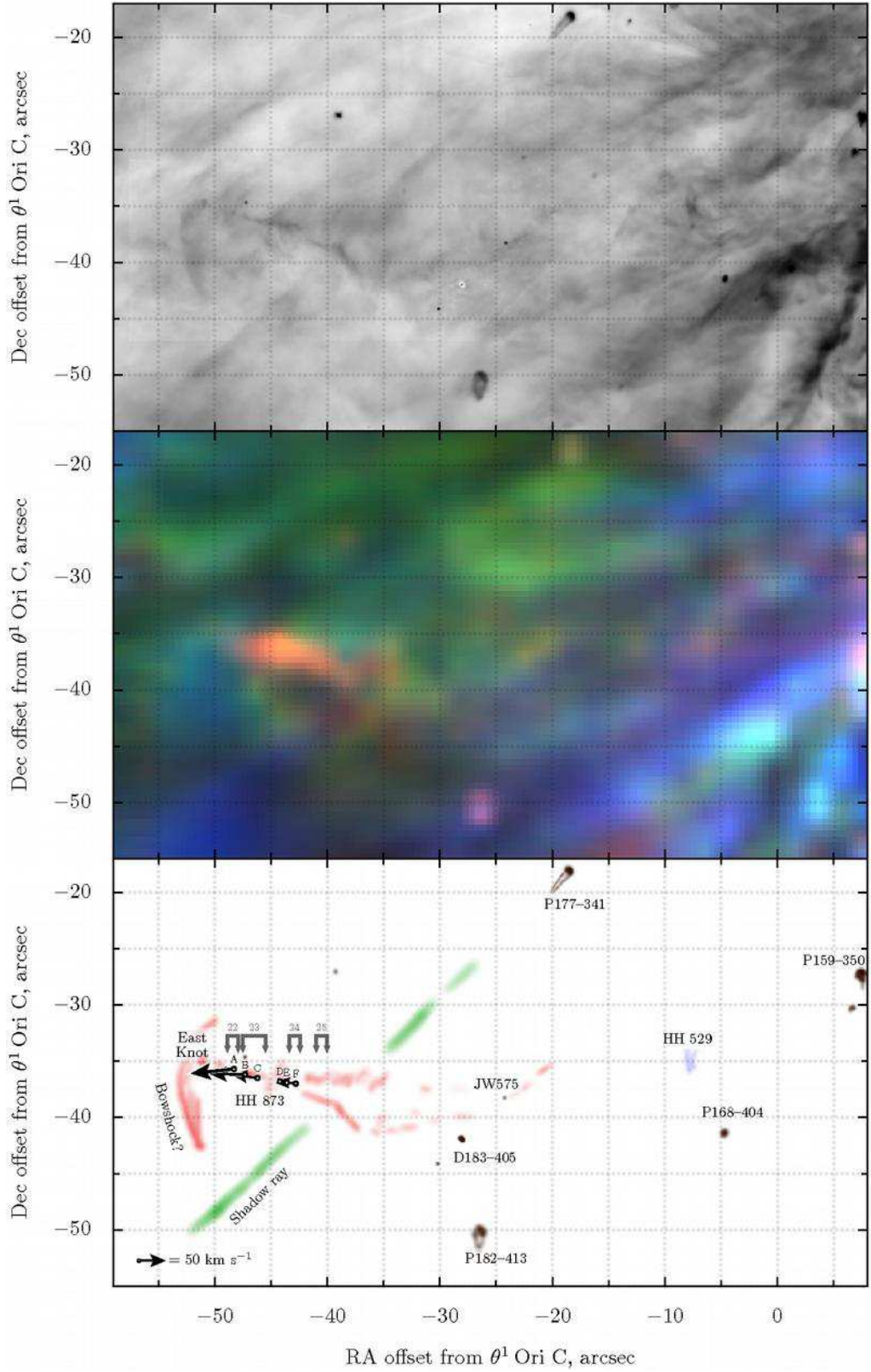}          
  \caption{Observations of the \HHnew{} redshifted jet. \textit{Top:}
    Portion of the \textit{HST} WFPC2 image in the \NII{}
    line. \textit{Center:} Three-channel color-coded \NII{} velocity
    map from ground-based observations, with $12~\kms$ wide channels
    centered on $\Vhel = 42$ (red), $30$ (green), and $18$ (blue)
    \kms. \textit{Bottom:} Finding chart for objects of interest in
    the image. Also shown are proper motions of individual knots
    (A--F) in the \HHnew{} jet and the positions of the slits (22--25)
    used to extract radial velocities.}
  \label{fig:hhnew}
\end{figure*}

This newly detected redsfifted jet is shown in greater detail in
Figure~\ref{fig:hhnew}. The top panel shows an \textit{HST} WFPC2
image of the region in the \NII{} line (f658n filter) at a resolution
of $\simeq 0.1\arcsec$, which is roughly 10 times better than the
resolution of our spectral observations. The jet is resolved into an
irregular, knotty structure, with long axis at $\mathrm{PA} \simeq
70\arcdeg$. The jet body is crossed prependicularly by several faint
arcs along its length, and its eastern end is marked by a larger
asymmetric arc, reminiscent of a bowshock.

From the morphology of the jet alone, it is not clear whether it is
travelling in an eastward or westward direction. If the large arc is
indeed a bowshock, then it would argue for an eastward motion. On the
other hand, a prominent knot at the eastern end of the jet (labelled
East Knot in the figure) bears some resemblance to a proplyd cusp,
which might represent the exciting source of the jet, arguing for a
westward motion, although this East Knot is not detected as a stellar
source in the near infrared \citep{2000ApJ...540..236H}, implying that
it is not in fact a proplyd.

\begin{table}
  \centering
  \caption{Proper motions of knots in the \HHnew{} jet}
  \label{tab:proper-motions}
  \setlength\tabcolsep{2\tabcolsep}
  \begin{tabular}{cccc}\toprule
         & Position & PA & Velocity, \Vtan \\ 
    Knot & (\arcsec) & (\arcdeg) & (\kms) \\ \midrule
    A & \xy{-48.3}{-35.9} & 96 & 81 \\ 
    B & \xy{-47.3}{-36.4} & 88 & 68 \\ 
    C & \xy{-46.2}{-36.7} & 82 & 48 \\ 
    D & \xy{-44.2}{-37.0} & 29 & 07 \\ 
    E & \xy{-43.6}{-37.1} & 92 & 20 \\ 
    F & \xy{-42.8}{-37.2} & 83 & 36 \\ \addlinespace[\smallskipamount]
    Error & & & $\pm10$ \\
    \bottomrule
  \end{tabular}
  \medskip
\end{table}

\begin{table}
  \centering
  \caption{Radial velocities and electron densities in the \HHnew{} jet}
  \label{tab:hh873-slits}
  \setlength\tabcolsep{2\tabcolsep}
  \begin{tabular}{cccc}\toprule
    & $x$ Position & Velocity, \Vhel & Density, \eden \\
    Slit & (\arcsec) & (\kms) & (\pcc) \\ \midrule
    22 & $-48.4$ & 37.2 & 3000 \\
    23 & $-46.5$ & 39.4 & 2500 \\
    24 & $-42.9$ & 38.0 & 3900 \\
    25 & $-40.5$ & 35.4 & 6200 \\ \addlinespace[\smallskipamount]
    Error & & $\pm 1.0$ & $\pm 500$ \\
    \bottomrule
  \end{tabular}
\end{table}

In order to resolve this issue, the proper motions of individual knots
along the jet have been determined (O'Dell, priv.~comm.) from a pair
of f658n images (8121SE and 5085S1 in Fig.~1 of
\citealt{2002AJ....124..445D}), separated in time by 5.19~years. The
results are given in Table~\ref{tab:proper-motions} and are shown
graphically in the lower panel of Figure~\ref{fig:hhnew}. These show
conclusively that the jet is moving towards the east, with the fastest
knot having a speed in the plane of the sky of $\simeq 80~\kms$ (the
errors in the proper motion measurements are of order $\pm 10~\kms$,
\citealt{2002AJ....124..445D}). Furthermore, the speed of the knots
shows a general increase from west to east. 

The contrast of the jet against the background nebula is highest in
the \OI{} line (see Fig.~\ref{fig:bispectra}\textit{c}) and it is also
detected in \SII{}, but it is not seen in \SIII{}. It is also clearly
visible in the \NIIlam{} line observations of DOH04, although it is
not commented on by those authors. The middle panel of
Figure~\ref{fig:hhnew} shows a color-coded isovelocity map of the
redshifted side of the \NII{} line from DOH04. The jet emission is
relatively brighter in the $42~\kms$ band and therefore shows up as
red in the image, against the blue/green of the surrounding nebula. 

We have measured radial velocities and electron densities in the jet,
using 4 of our \SII{} slit spectra that cross it. The positions and
widths of the slits are indicated by gray arrows in the bottom panel
of Figure~\ref{fig:hhnew}, and the results are given in
Table~\ref{tab:hh873-slits}. For each slit, the line profile in a
2\arcsec{} long section across the jet was fitted by 3 gaussians,
which represent the nebula, the jet, and the Diffuse Blue
Layer. Essentially identical results for the radial velocities are
obtained from \NII{} and \OI{}. 

The electron density of the background nebula in the same slits ranges
from $1500$ to $2000~\pcc$ and it can be seen from
Table~\ref{tab:hh873-slits} that the jet is significantly denser than
this, especially at its western end. There is evidence for an increase
in the radial velocity as one passes along the jet towards the east,
except for the easternmost slit (\#22), where the velocity decreases
again. The jet component is also weaker by a factor of two in this
slit, compared with the others, which is consistent with the
brightness of the jet as seen in the \textit{HST} image. In the
westernmost slit (\#25), the jet component becomes more spatially
extended, corresponding with its broader appearance on the
\textit{HST} image at that position. No detectable redshifted
component was found in slits that cross the East Knot or the bowshock.
The jet component in all the spectra was found to have an intrinsic
width (FWHM) of $\simeq 9~\kms$ after correcting for instrumental
broadening.

It is impossible to compare individual slit positions with individual
proper motion knots due to the wide disparity in the spatial
resolution. Instead, we compare only the average radial and tangential
velocities of the jet. We eliminate proper motion knot D from this
average because its anomalously low velocity would cause its emission
to be hidden inside the nebular background component in our
spectra. The average tangential velocity is thus $\Vtan = 51 \pm
12~\kms$ in direction $\mathrm{PA} = 89 \pm 3\arcdeg$. To calculate
the mean radial velocity, it is necessary to subtract the velocity of
the stellar cluster, $\Vhel = 28 \pm 2~\kms$
\citep{1988AJ.....95.1755J}, giving $V_\mathrm{r} = 9.5 \pm 2~\kms$.

Since the jet is not exactly straight, it is hard to say with any
certainty what its exciting source may be, although some possible
candidates are labelled in the lower panel of
Figure~\ref{fig:hhnew}. The dark silhouette proplyd disk
\OW{D183}{405} lies close to the western end of the long continous
section of jet, but the position angle of the elliptical disk
silhouette is not perpendicular to the jet axis, as would be expected
on theoretical grounds. Besides, there is some indication that the jet
extends further to the west than \OW{D183}{405}, passing close to the
star JW575, which is another possible source. However, the jet may
extend much farther to the west even than this, as evidenced by a
broken sequence of faint red knots in the middle panel of
Figure~\ref{fig:hhnew}, which terminates in the general area of the
bright proplyd \OW{P159}{350}. This proplyd itself is unlikely to be
the source, since it is known to have a jet pointing to the NNE
\citep{2000AJ....119.2919B}, but it also has a fainter binary
companion which might be the source. Unfortunately, any attempt to
trace the jet over large distances is hampered by the presence of many
other high speed flows in the same general area, such as \HH{529}.

\subsubsection{South Blue Arc}
\label{sec:south-blue-arc}

Starting roughly 10\arcsec{} south of the nose of the \HH{204} bowshock,
we detect a very faint arc of blueshifted \SII{} emission, which runs
about 30\arcsec{} to the south, while bending slightly to the west
(see bottom-left panel of Fig.~\ref{fig:channel-broad-blue}). This arc
is most visible around $-10~\kms$ and is distinguished from the
emission of the diffuse blue layer, both by its slightly bluer
velocity and by its density of $\simeq 1000~\pcc{}$, which is roughly
twice that of the diffuse blue layer, as can be seen in the top-left
panel of Fig.~\ref{fig:ratio-a}).

This blueshifted emission arc may possibly represent a third bowshock
due to the same flow that gives rise to \HH{203/204}. It is probably
associated with a cocoon of faint, high-ionization emission detected
in deep \textit{HST} imaging of the same region
\citep[\S~3.5]{Henney-OrionSouth}.  DOH04 present arguments against
the possibility of a common origin for \HH{203} and \HH{204} but these
are not conclusive. We suggest that the feature shown in Fig.~10 of
DOH04, rather than being a second jet that drives \HH{204}, might
instead be associated with the \HH{203} jet. In particular, the knots
at \OW{204}{448} and \OW{199}{447} seem to form part of one side of
the nose of a fourth bowshock, with the other side formed by the
feature that branches up at $\mathrm{PA} \simeq 330\arcdeg$ from
\OW{204}{448}. This could either be an internal working surface or due
to an interaction of the jet with the Big Arc. It is worth noting that
the highly blueshifted emission from the \HH{203} jet does show a
bright knot at this position (\OW{198}{441} in Fig.~9 of DOH),
together with several abrupt changes in direction along its length,
tending towards smaller position angles as one approaches the source
of the jet. The progression of radial velocities towards greater
blueshifts from the South Blue Arc, through \HH{204}, to \HH{203}
suggests that the jet direction is also moving closer to the line of
sight. Such structures are similar to those seen in simulations of
episodic, precessing jets \citep{2001MNRAS.327..507L,
  2004A&A...426L..25C}.

\subsection{Proplyds}
\label{sec:proplyds}

\begin{table}
  \centering
  \caption{Proplyds identified in our spectra}
  \label{tab:proplyd}
  \setlength\tabcolsep{2\tabcolsep}
  \begin{tabular}{ccc}
    \toprule
    & Position & $V_\mathrm{peak}$(\OI{}) \\
    ID & (\arcsec{}) & (\kms{}) \\
    \midrule
    P155--338 & \xy{+14}{-14} & +23\\
    P159--350 & \xy{+07}{-26} & +23\\
    P167--317 & \xy{-03}{+06} & +30\\
    P168--326 & \xy{-05}{-02} & +31\\
    P170--337 & \xy{-08}{-13} & +30\\
    P177--341 & \xy{-18}{-17} & +25\\
    P180--331 & \xy{-22}{-07} & +26\\
    P182--413 & \xy{-25}{-49} & +25\\
    P206--446 & \xy{-61}{-82} & +20\\
    \bottomrule
  \end{tabular}
  \medskip
\end{table}

We can identify several of the bright proplyds in our data, which are
listed in Table~\ref{tab:proplyd}, together with the peak \OI{}
velocity of each proplyd, as determined from our spectra. The proplyds
are best seen in \OI{} since the high electron densities (up to
$10^6~\pcc$, \citealp{2002ApJ...566..315H}) seen in these objects
imply that the \SII{} and \SIII{} lines suffer from collisional
deexcitation. They show up well in the \OI{}/\SII{} ratio,
particularly on the blue side of the line. A few of the proplyds are
detected in \SIII{}, where they tend to show greater line widths than
the surrounding nebula, but none are convincingly detected in \SII{}.
The positions and velocities of these proplyds are also shown in
Figure~\ref{fig:finding}\textit{e}. Due to confusion and the effects
of scattered light, the proplyds close to the Trapezium are harder to
spot in our data, particularly those to the west, and as a result some
of the brightest proplyds (e.g., LV~5) are missing from our sample.

\citet{1999AJ....118.2350H} showed that, although the \OI{} line can
show blue or red wings due to the photoevaporation flow, the peak
\OI{} velocity is indicative of the radial velocity of the proplyd's
central star. For our sample of proplyds we find a mean velocity of
$\Vhel = 26~\kms$ and a one-dimensional velocity dispersion (after
correcting for an estimated 2~\kms{} measurement error) of $\sigma =
2.9~\kms{}$. This latter is not significantly different from the value
of $2.5~\kms$ obtained from proper motion studies of the star cluster
\citep{1988AJ.....95.1755J}. The mean blueshift of the proplyd stars
with respect to the background molecular gas (at $\Vhel =
27$--29~\kms{}) is much less than was found by
\citet{1999AJ....118.2350H} for a smaller sample. To date, only 6
proplyds have reliably measured inclination angles, which require
fitting kinematic models to multiline spectroscopy
\citep{1999AJ....118.2350H, 2002ApJ...566..315H,
  2002ApJ...570..222G}. By combining our data with the \NII{},
\OIII{}, and H$\alpha$ datasets of DOH04 it should be possible to
extend this technique to a much larger sample and hence determine the
three-dimensional distribution of proplyds in the nebula.

\subsection{The Red Bay}
\label{sec:red-bay}

This is a large region of relatively redshifted \SIII{} emission, most
prominent in the $+25$ to $+35~\kms$ range (red emission in
Fig.~\ref{fig:channel-broad} and blue emission in
Fig.~\ref{fig:channel-broad-red}), which extends out to the east and
east-southeast from the Trapezium. On the blue side of the line
profile ($< +20~\kms$), the Bay region is almost devoid of \SIII{}
emission (see lower-right panel of Fig.~\ref{fig:channel-broad-blue}),
whereas on the far-red side it also corresponds to a minimum in the
brightness of the scattered light (\S~\ref{sec:red-scatt-comp}). All
these features are illustrated in Figure~\ref{fig:finding}\textit{d}.

In this roughly elliptical, $120\times80\arcsec$ region, the blueshift
between the high-ionization lines and low-ionization lines is reduced
to $\simeq 2~\kms$, as compared to the $\simeq 10~\kms$ that is more
typical in the west of the nebula. The fact that the Red Bay has the
appearance of an empty cavity in blueshifted velocity channels might
be taken to imply that the region could be explained by the action of
the stellar wind from the Trapezium stars in confining the ionized gas
to a thin layer near the ionization front. However, despite its
appearance in position-velocity space, the Bay is \emph{not} a thin
layer but rather has an effective thickness along the line of
sight\footnote{The effective thickness \citep{2005ApJ...627..813H} is
  calculated by dividing the emission measure (calculated from the
  radio free-free brightness) by the electron density (measured from
  our \SII{} observations).} of $\sim 10^{18}~\mathrm{cm}$, which is
comparable to its lateral extent on the sky.

We therefore propose that the Red Bay is a relatively inert region of
the nebula, which partakes only weakly in the general champagne
flow. Maps in molecular lines and far-infrared dust emission
\citep{1999ApJ...510L..49J} show that there is relatively little dense
molecular gas behind the nebula at this position, which may have
permitted the ionization front to propagate further along the line of
sight away from the observer. This region would therefore be bounded
from behind by a concave, weak-D ionization front, the bowl-like shape
of which would be expected to trap an extended zone of slow-moving
ionized gas (see, for example, \citealp{2006ApJ...647..397M}).

\subsection{The red scattered component }
\label{sec:red-scatt-comp}

In the far-red wing ($> +45~\kms$) of the lines we detect a pattern of
bright and dark lanes with a morphology that is almost identical
between the 3 different lines. It can be seen as the orange/red
emission in Figure~\ref{fig:channel-broad-red} and is more prominent
in \SII{} and \SIII{} than in \OI{}. This probably represents nebular
emission that has been scattered by dust in the PDR and molecular
cloud behind the ionization front. Because the emitting gas is moving
away from the dust in the background cloud, the scattered emission
will be redshifted \citep{1987A&A...186..322L, 1992ApJ...399L..67O,
  1998ApJ...503..760H}. Figure \ref{fig:finding}\textit{d} shows the
general layout of the more prominent lanes and also indicates (dotted
line) the boundary of an extended plateau of far-red emission, which
is found just to the east of the Trapezium. The relationship between
the scattered component and the bright bars and dark lanes is shown in
Figure~\ref{fig:finding}\textit{f}.

\subsection{The Southeast Diffuse Blue Layer }
\label{sec:diffuse-blue-layer}

\begin{figure}
  \centering
  \includegraphics{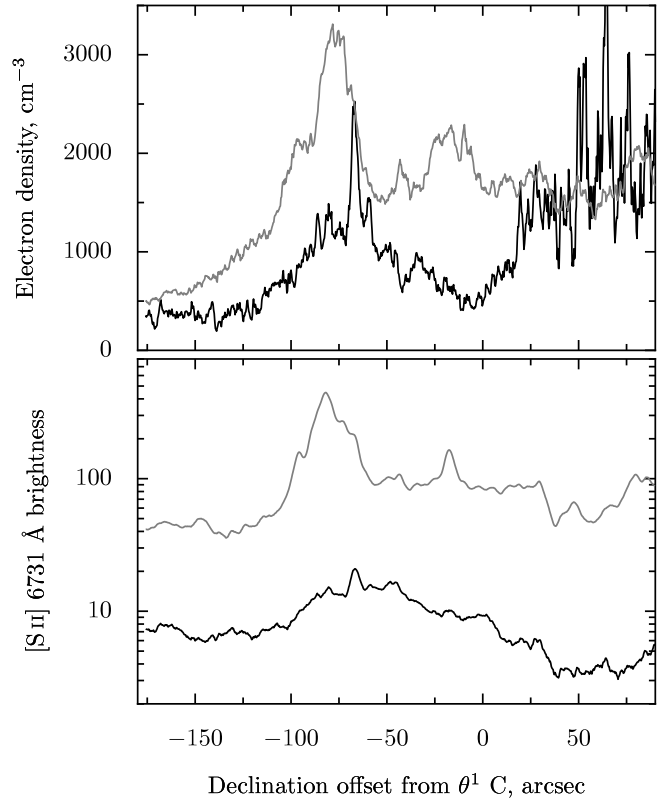}          
  \caption{North-South profiles of electron density (upper panel) and
    \SII{} surface brightness (lower panel) from a slit situated
    $40.5\arcsec{}$ E of \thC{}. The black lines show emission at
    velocities around that of the blueshifted line component ($-7$ to
    $+7$~\kms{}) while the gray lines show emission from the principal
    line component ($+7$ to $+50~\kms{}$).}
  \label{fig:bluelayer}
\end{figure}

A notable property of the \SII{} profiles in the southeastern region
of the nebula (Fig.~\ref{fig:bispectra}\textit{cd}) is that they are
double-peaked, showing a well-defined secondary emission component
centered on $\Vhel \simeq -2~\kms$, which is 20--25~\kms{} to the blue
of the main component. This component is not detected in \OI{} or
\SIII{}, although to the north of the Bright Bar it merges with
emission from the Big Arc, which is also visible in \SIII{}. The
spatial distribution of this component, which we will call the Diffuse
Blue Layer, can be appreciated as the extended green emission in the
lower left corner of the \SII{} image of
Figure~\ref{fig:channel-broad-blue}. The electron density of this
layer is very low, as can be seen from Figure~\ref{fig:ratio-a}.

In the lower panel of Figure~\ref{fig:bluelayer} we show brightness
profiles for red and blue velocity ranges of the \SIIlam{} line for a
representative slit in the East of the nebula. The blue channel covers
velocities $-7$ to $+7$~\kms{} while the red channel covers velocities
$+7$ to $+50~\kms{}$. The Diffuse Blue Layer is the dominant
contribution to the blue channel brightness south of an offset of
$-100$. The upper panel shows that the density of the Layer is roughly
constant at $\eden \simeq 400~\pcc$. The Bright Bar crosses the slit
for positions between $-75$ and $-100$ and the brightness of the
systemic nebular component becomes so large that its wing also
dominates the brightness in the blue channel. Slightly further to the
north, a peak in blue channel brightness occurs where the Big Arc
crosses the slit. Therefore, the density rise in the blue channel that
is seen between positions of $-100$ and $-25$ is not due to the
Diffuse Blue Layer itself. To the north of the Big Arc, between
offsets of $-25$ and $0$, the Diffuse Blue Layer again dominates the
blue channel and the density returns to $\simeq 500~\pcc$. Further to
the north, the intensity of the Diffuse Blue Layer declines and the
wing of the systemic component returns to dominate the blue channel.

To the west of the nebula, the Diffuse Blue Layer is much less
prominent and is only detectable to the south of the Bright Bar.  A
dark band appears to cross the Diffuse Blue Layer in the $-12~\kms{}$
channel, roughly parallel to the ionization front of the Bright Bar
but about 15\arcsec{} farther away from the Trapezium.
 
One of our \SII{} slits reaches 40\arcsec{} farther north than the
area covered by our maps and shows that a similar feature to the
Diffuse Blue Layer reappears around \xy{+49}{+156}. The density is
$\eden \simeq 500~\pcc$ but the velocity is slightly redder at $\Vhel
\simeq +9~\kms{}$, although it is still well separated from the
systemic component, which lies at $+24~\kms{}$ with $\eden \simeq
1300~\pcc$.

We do not detect this component in \OI{} or \SIII{} but it can also be
seen in \NII{} \citep{1999AJ....118.2350H} and in [\ion{O}{2}]~3727,
3729~\AA{} \citep{1992PhDT........35J}. The
\citeauthor{1999AJ....118.2350H} observations show it to extend at
least another 20\arcsec{} east of the region covered by our maps. It
seems likely that this component is the same as that identified many
years ago as a region of line splitting in \NII{}
\citep{1973A&A....29..341D}. The region around and to the south of the
Bright Bar coincides with \citeauthor{1973A&A....29..341D}'s Region~A,
while the similar feature found to the North is part of her
Region~B. A third region of splitting (Region~C) in the far west of
the nebula lies outside the area of our observations.

The lack of \OI{} emission suggests that this emission does not come
from an ionization front, but rather from an extended region of fully
ionized gas. The lack of \SIII{} emission implies that helium is
neutral and that the ionizing spectrum is rather soft. It is therefore
possible that the Blue Layer is due to a separate region along the
line of sight that is ionized by \thA{} instead of by \thC{}. This is
also consistent with the spatial distribution of the layer in the
south east (\citeauthor{1973A&A....29..341D}'s Region A). However, the
larger scale distribution of the \NII{} line splitting mapped by
\citet{1973A&A....29..341D} argues instead that the Blue Layer is
instead related to the large-scale champagne flow from the
nebula.

\section{Conclusions}
\label{sec:conclusions}

We have mapped the Orion Nebula in a variety of optical emission lines
in order to produce images of the region that are resolved in both
velocity and in ionization. We have also produced the first
velocity-resolved electron density maps of the nebula. The
spatio-kinematic information revealed by our maps sheds new light on
previously known nebular features and also uncovers several new
components of the nebula. In this paper, we concentrate on low
ionization emission within $\pm 40~\kms$ of the systemic velocity and
our main conclusions are:
\begin{enumerate}
\item We find a region to the east and southeast of the Trapezium,
  which we call the Red Bay, where the usual correlation between
  velocity and ionization potential is very weak or absent and where
  the emission layer is much thicker along the line of sight than in
  the west of the nebula. This is gas that is not sharing in the
  general champagne flow of the nebula, presumably beacause of the
  local concave geometry of the ionization front in the region.
\item We find an extensive layer of low-density, low-ionization,
  blueshifted emission in the southeast of the nebula, which may be
  ionized by the softer spectrum of the nearby O9.5 star \thA, or may
  represent the near side of the champagne flow from the nebula.
\item We detect a new redshifted jet, \HHnew{}, to the southwest of
  the Trapezium, and a dense low-ionization shell that may represent
  an outer bowshock in the \HH{203/204} flow.  
\end{enumerate}

\acknowledgments

We are grateful to Bob O'Dell and Takao Doi for allowing us to use
their KPNO \SII{} data, to Bob O'Dell for carrying out the proper
motion measurements, to Michael Richer and Bob O'Dell for assistance
during some of the SPM observations, and to Jane Arthur, Bob O'Dell,
and John Meaburn for useful discussions. Alex Raga suggested the use
of the de-jittering technique described in
\S~\ref{sec:observations}. We thank Bo Reipurth for assigning a number
to \HHnew. This research has made use of the observational facilities
at San Pedro M\'artir Observatory, B.C., Mexico, and Kitt Peak National
Observatory; NASA's Astrophysics Data System Bibliographic Services;
arXiv.org e-print archive, operated by Cornell University; the SIMBAD
database, operated by the Centre de Donn\'ees astronomiques, Strasbourg;
and SAOImage DS9 \citep{2003ASPC..295..489J}, developed by the
Smithsonian Astrophysical Observatory. We acknowledge financial
support from DGAPA-UNAM, Mexico, through project IN115202, and
CONACyT, Mexico, through a research studentship to MTGD.

\appendix

\section{Translation between the coordinates used in the current paper
  and those of O'Dell \& Wen (1994)}

In the literature it is common for objects in the Orion nebula to be
designated by a six-digit coordinate-based code of the form
\OW{XXX}{YZZ}, which was first introduced by
\citet{1994ApJ...436..194O} and which is based on a truncated form of
their J2000 positions, rounded to $0.1$~seconds in RA and $1\arcsec$
in declination (e.g., DOH04). For example, in this system, \thC{}
would have the designation \OW{165}{323}. In the current paper, we do
not use this system, preferring instead to simply give arcsecond
offsets \xy{x}{y} with respect to \thC{}. 

In order to ease the comparison of our results with those of other
papers, we here provide equations to transform between the two
coordinate systems. The transformation between our \xy{x}{y} offsets
and \citeauthor{1994ApJ...436..194O}'s system is
\begin{eqnarray}
  \mathrm{XXX} & = & \lfloor 164.6 - 0.6696\,x \rfloor \nonumber \\
  \mathrm{Y}   & = & \lfloor (203.2 - y)/60 \rfloor \nonumber \\
  \mathrm{ZZ}  & = &\lfloor 203.2 - y \rfloor - 60\,\mathrm{Y} , \nonumber
\end{eqnarray}
in which the notation $\lfloor \cdots \rfloor$ denotes the integer
part, padded with leading zeros to the requisite number of digits.
The inverse transformation is
\begin{eqnarray}
  x & = & 245.8 - 1.493\,\mathrm{XXX} \nonumber \\
  y & = & 203.2 - (60\,\mathrm{Y} + \mathrm{ZZ}). \nonumber
\end{eqnarray}
Note that these equations are only valid for right ascensions between
05\hour{} 35\minute{} and 05\hour{} 36\minute{} and declinations south
of $-05\arcdeg{}\ 20\arcmin{}$, which corresponds to $-650 < x < 246$ and
$y < 203$. All our observations fall well within these limits.


\end{document}